\documentclass[12pt, a4paper, final]{article}
\usepackage{amsmath}
\usepackage{amssymb}
\usepackage[english]{babel}
\usepackage{authblk}
\usepackage[colorlinks=true, linktocpage=false, draft=false]{hyperref}
\usepackage[doipre={doi:~}, citeurlpre={url:~}, citeurlpost={}]{uri}
\usepackage[left=3cm, right=1.5cm, top=2cm, bottom=2cm]{geometry}
\usepackage[style=gost-numeric-min, blockpunct=space, language=auto, autolang=other, sorting=none, bibdoi=true, biburl=true, movenames=false, maxnames=4, minnames=3, useeditor=true, backend=biber]{biblatex}

\AtEveryBibitem{\clearfield{month}}
\AtEveryBibitem{\clearfield{number}}
\AtEveryBibitem{\clearfield{issue}}
\DeclareFieldFormat[article]{title}{}
\DeclareFieldFormat{doi}{%
  doi\addcolon\space
  \ifhyperref
    {\href{https://doi.org/#1}{\nolinkurl{#1}}}
    {\nolinkurl{#1}}}
\DeclareFieldFormat{url}{%
  url\addcolon\space
  \ifhyperref
    {\href{#1}{\nolinkurl{#1}}}
    {\nolinkurl{#1}}}

\allowdisplaybreaks

\makeatletter
\predisplaypenalty=\@medpenalty
\makeatother

\title{Torsional Alfven Oscillation in the Regime of Firehose Instability as a Mechanism of Plasma Stratification in a Laboratory Experiment on Modeling a Coronal Arch}

\author[1,2]{S.~A.~Koryagin}
\author[1,2]{M.~E.~Viktorov}
\affil[1]{A.~V.~Gaponov-Grekhov Institute of Applied Physics of the Russian Academy of Sciences, Nizhny Novgorod, 603950 Russia}
\affil[2]{Lobachevsky University, Nizhny Novgorod, 603022 Russia}

\begin{document}
\linespread{2}

\maketitle

\abstract{The compact laboratory stand ``Solar Wind'' (Inst. Appl. Phys. of Russ. Acad. Sci.) forms an arch structure of the coronal loop type, in which the plasma pressure varies from zero to values of the order of and above the magnetic pressure. The arc discharge in each of the bases of the magnetic tube creates a plasma that is characterized by a significantly higher ion temperature along the magnetic field line than across the latter one. In the~stationary state (when the ion pressure is below the threshold value for rupture of the system at the top of the loop), the plasma is found to be stratified in the form of a cylindrical layer along the outer wall of the tube or possibly two belts along the upper and lower vaults of the arch. The paper discusses the excitation of a torsional Alfven oscillation in the loop in the regime of firehose instability. In~the case of rapid growth (with an increment of the order of an ion cyclotron period), the unstable Alfven oscillation essentially reallocates the particles between the central axis and the tube wall, which manifests itself in the form of the observed cylindrical layer.}

\subsection*{INTRODUCTION}

Magnetic loops are the basic structural unit of the solar corona and contain a significant mass of the latter \cite{Zaitsev-2008-transl}. Their length varies from the height of the homogeneous chromosphere to a value of the order of the solar radius. The characteristic tube diameter is determined by the spatial scale of photospheric convection. Single loops are able to unite into ribbon arcades and, on the contrary, to stratify into a bundle of thinner fibers due to the flute instability.

The small tube radius $r$ compared to its length $L$ (and bending radius) indicates a low ratio $\beta$ of the plasma and magnetic pressures, since the flute instability limits the parameter $\beta$ to a value on the order of the ratio $r/L$ \cite[\S~11.5]{Mikhailovsky-II-book-transl}. In eruptive prominences, the plasma $\beta$ appears to approach unity at the point of structure rupture.

Magnetic loops form resonators (waveguides) for magnetohydrodynamic waves: Alfven, magneto-sonic, and ion-sonic \cite{Zaitsev-book-eng} waves. The damping of the waves is due to both their emission into the environment and collisional heating of the particles. Depending on their polarization, Alfven and magneto-sonic oscillations take the form of tube bends (kinks) or torsional oscillations. In the latter, the torsion of magnetic force lines occurs along the azimuth inside the plasma fiber \cite{Kohutova-2020}.

In turn, the solar wind is characterized by a higher plasma pressure compared to the magnetic pressure, as well as by anisotropic ion temperature relative to the direction of the local magnetic field (due to the different degree of plasma expansion along and across the magnetic force line). The firehose instability of Alfven and magnetosonic wave \cite{Yoon-2019} is considered as a possible mechanism of ion temperature isotropization. At the same time, the firehose instability of the plasma in the tail of the Earth's magnetosphere has been theoretically studied as a source of irregular low-frequency pulsations of the Pi2 type during substorm \cite{Wong-2005}.

The experimental setup ``Solar Wind'' (IAP RAS) was created to model a plasma structure of the single coronal tube type: to elucidate the conditions of magnetic confinement and destruction of the system in the form of a coronal mass ejection with generation of nonthermal radio emission \cite{Viktorov-2015-transl, Viktorov-2017}. The plasma configuration is characterized by a significantly higher ion temperature along the magnetic tube axis than in the transverse direction. In turn, the sufficiently large diameter of the plasma cord with respect to its length and bending radius allows to confine the plasma with relatively high transverse pressure without triggering flute instability. Under these conditions, the longitudinal pressure of the plasma is capable of exceeding the magnetic pressure and reaching the threshold level for the development of the firehose instability. The purpose of the present work~--- to elucidate the mechanism of plasma stratification in the experimental setup, which occurs without destroying the system.

Section~1 summarizes the basic parameters of the experimental setup of the ``Solar Wind''. Section~2 discusses the limit of collisionless ion motion at the length of the magnetic tube at which the temperature anisotropy of the ion fraction~--- a necessary factor for firehose instability~--- is reached. In~section~3, the conditions under which plasma stratification is detected in the experiment are given; the attainment of the threshold of firehose instability at the top of the magnetic arch is justified. Section~4 draws attention to the coincidence of the characteristic spatial scale of stratification in the setup and in known self-consistent current structures \cite{Veselovsky-1975-transl, Kocharovsky-2016-transl} in plasmas with temperature anisotropy. Section~5 outlines a physical model of the torsional Alfven oscillation in an inhomogeneous magnetic tube, which qualitatively follows the approach \cite{Wong-2005} but is adapted for subsequent analytical rather than numerical study. 

In contrast to a homogeneous medium, an inhomogeneous system requires a special condition of coupling of magnetohydrodynamic solutions in the boundary section between stable and unstable regions, on which a thin current layer (with a thickness of the order of the ion gyroradius) of the type of a domain wall in an antiferromagnet is formed. This circumstance requires a kinetic description of ion motion in the system. In~turn, the model spatial profile of the magnetic field reduced the problem to the classical Sturm~--- Liouville problem of mathematical physics for the Legendre equation, but in this problem the zero boundary conditions occur at the ends of the segment containing the regular special points of the differential equation. In this case, the discrete spectrum of eigenvalues is unbounded from below in magnitude, which is characteristic of the so-called left-defined Sturm~--- Liouville problem \cite{Mingarelli-2011, Richardson-1918}, while in this right-defined problem it comes from the change of sign of the coefficient at the senior derivative in the equation \cite{Atkinson-1987}.

The rather compact size of the experimental system in units of the ion Larmor radius allows only a small number of unstable torsional modes and prevents the development of a turbulent cascade in contrast to an unbounded medium. At the same time, this circumstance provides an increment of firehose instability at the level of the ion cyclotron frequency, what is necessary for a significant perturbation of the plasma density in Alfven motion. The above arguments substantiate the possibility of the observed plasma layering due to the firehose instability and are discussed in detail in section~6 and the concluding part of the paper.

\subsection*{1.~EXPERIMENTAL STAND ``SOLAR WIND''}

The compact experimental stand ``Solar Wind'' (IAP RAS) allows one to create a plasma system similar to a coronal loop \cite{Viktorov-2015-transl, Viktorov-2017}. Powerful solenoids encompass two mutually perpendicular radial entries inside the vacuum chamber (the latter has a diameter of $2R=15~\text{cm}$) and form a curved arch-shaped magnetic tube. The arch looks like a quarter circle with length $L_\text{loop}=\pi R/2=12~\text{cm}$. The diameter and length of each solenoid $l_\text{sol}$ is approximately equal to the diameter $63~\text{mm}$ of the flanged chamber inlet. On the outer (with respect to the chamber inlet) slice of each solenoid is placed a plasma source in the form of a localized arc discharge on an aluminum disk $10~\text{mm}$ in diameter. A $20~\text{mm}$ diameter steel electrode covers the perimeter of the discharge and plays the role of a well-conducting base of the plasma cord. 

In~the absence of a magnetic field, each plasma generator releases an almost completely ionized supersonic plasma stream with hydrodynamic velocity $v_{\text{i}\parallel}=15~\text{km}/\text{s}$. The latter corresponds to the Mach number $M_\text{s}=3{.}5$ by the ion-sonic velocity $c_\text{s}=\sqrt{ZT_\text{e}/m_\text{i}}=4{, }3~\text{km}/\text{s}$ for electron temperature $T_\text{e}=3~\text{eV}$, mean ion charge number $Z=1{.}7$, and ion mass $m_\text{i}=27~\text{u}$ Encountering supersonic flows is a system within which the kinetic energy of the relative motion of ions along the magnetic field significantly exceeds their thermal energy in the orthogonal plane. The anisotropy of the ion temperature gives rise to the firehose instability. The latter is a magnetic field-modified ionic Weibel instability \cite{Mikhailovsky-I-book-transl}.

\subsection*{2.~COLLISIONLESS LIMIT FOR IONS AND~ANISOTROPY OF THEIR TEMPERATURE}

After passing inside the solenoid, the diameter of the plasma cord at the chamber entrance $2r_\text{foot}=10~\text{mm}$ coincides with the diameter of the aluminum cathode. Therefore, the ionic density at the base of the loop $n_\text{i\,foot}$ (in each of the counter streams) is approximately equal to its value in the plasma generator \cite{Viktorov-2017}:
\begin{equation}
n_\text{i\,foot}=2\cdot10^{14}\*\left(I_\text{gen}/1~\text{kA}\right)~\text{cm}^{-3}\,.
\label{IonDensityInArcDischarge}
\end{equation}
At the maximum discharge current $I_\text{gen}^\text{(max)}=8~\text{kA}$, the quantity (\ref{IonDensityInArcDischarge}) reaches a magnitude
\begin{equation}
n_\text{i\,foot}^\text{(max)}=1{.}6\cdot10^{15}~\text{cm}^{-3}.
\label{MaximalIonDensityInOneFlowAtArcFoot}
\end{equation}
On the way from the base to the top of the loop (the apex of the arch is located almost in the center of the vacuum chamber), the diameter of the plasma cord expands $\sqrt{\alpha}=3$~fold. This thickening of the tube corresponds to a $\alpha=9$~fold decrease in magnetic induction and plasma density in the flow along the same path (assuming collisionless ion motion).

Coulomb collisions between counter-flux ions transform their kinetic energy of motion along ($K_{\parallel}$) and across ($K_{\perp}$) the magnetic field at the rate $\mathrm{d}K_{\parallel}/\mathrm{d}t=-\mathrm{d}K_{\perp}/\mathrm{d}t=-2\nu_\text{tr\, ii}K_{\parallel}$, where the transport frequency $\nu_\text{tr\,ii}=4\pi n_\text{i}^{(1)}r_{\text{si}\parallel}^2\,(2v_{\text{i}\parallel})\,\ln(r_\text{D}/r_{\text{si}\parallel})$ characterizes the negative acceleration of the ion $\mathrm{d}v_{\text{i}\parallel}/\mathrm{d}t=-\nu_\text{tr\,ii}v_{\text{i}\parallel}$ along the line of force; $r_{\text{si}\parallel}=(Ze)^2/[(m_\text{i}/2)\,(2v_{\text{i}\parallel})^2]$~--- the impact parameter of the scattering at $90^\circ$ of two particles with relative velocity $2v_{\text{i}\parallel}$ and reduced mass $m_\text{i}/2$, $r_\text{D}=[T_\text{e}/(8\pi Zn_\text{i}^{(1)}e^2)]^{1/2}$~--- the electrostatic shielding radius of the ion by the electrons of both streams; $e$~--- elementary charge; $n_\text{i}^{(1)}$~--- ion density in one stream. The energy $K_{\parallel}$ decreases by a factor of $2{.}7$ over the length of the loop $L_\text{loop}$ (and the ionic temperature is isotropyzing) if the ionic density $n_\text{i}^{(1)}$ at the top of the arch exceeds the value of
\begin{multline}
n_\text{i\,col}^\text{(max)}=
\frac{1}{16\pi L_\text{loop}r_{\text{si}\parallel}^2\ln(r_\text{D}/r_{\text{si}\parallel})}
=\\=
1{.}5\cdot10^{13}
\left(\frac{L_\text{loop}}{12~\text{cm}}\right)^{-1}
\left(\frac{Z}{1{.}7}\right)^{-4}
\left(\frac{v_{\text{i}\parallel}}{1{.}5\cdot10^6~\text{cm}/\text{s}}\right)^4
\left[\frac{\ln(r_\text{D}/r_{\text{si}\parallel})}{10}\right]^{-1}~\text{cm}^{-3}.
\label{LowerIonDensityForTemperatureEquipartition}
\end{multline}

Highest achievable ionic density at the loop top in a single flow
\begin{equation}
n_\text{i\,top}^\text{(max)}=n_\text{i\,foot}^\text{(max)}/\alpha=1{.}8\cdot10^{14}~\text{cm}^{-3}
\label{MaximalIonDensityInOneFlowAtArcTop}
\end{equation}
exceeds the value (\ref{LowerIonDensityForTemperatureEquipartition}) by a factor of 10. This fact indicates the realization of both strongly collisional and collisionless regimes for the ion fraction in the experiment. We investigate a variant of the collisionless system.

\subsection*{3.~PLASMA LAYERING. FIREHOSE INSTABILITY}

Optical images of the system \cite{Viktorov-2019} reveal plasma layering at maximum discharge current ($8~\text{kA}$) and unperturbed magnetic induction at the loop apex 
\begin{equation}
B_\text{strat}=700~\text{G}.
\label{MagneticInductionOfObservedPlasmaLayering}
\end{equation}
The stratification appears as a brighter glow of the medium along the upper and lower vaults of the arch at lower ionic density at the cord's core axis and allows interpretation as: a)~two belts of increased density; b)~a thin near-wall cylindrical layer. The photo was taken from the ``facade'' of the arch and therefore does not allow to choose only one variant of layering.

In the case of plasma density (\ref{MaximalIonDensityInOneFlowAtArcTop}) and magnetic induction (\ref{MagneticInductionOfObservedPlasmaLayering}), the longitudinal pressure of the ions of the oncoming flows $P_{\text{pl}\parallel}=2n_\text{i\, top}^{(\text{max})}\,m_\text{i}v_{\text{i}\parallel}^2$ at the top of the loop exceeds the unperturbed magnetic pressure $P_\text{mag}=B_\text{strat}^2/(8\pi)$ almost exactly 2~fold:
\begin{equation*}
\beta_{\parallel}=\frac{P_{\text{pl}\parallel}}{P_\text{mag}}=1{.}9.
\end{equation*}
At a threshold value of $\beta_{\parallel}=2$, a firehose Alfven wave instability develops in the limit of strong temperature anisotropy of ions and cold electrons \cite[Chap.~3, \S~10]{Krall-book-eng}.

Under conditions of firehose instability ($\beta_{\parallel}>2$), the Alfven wave transforms into an aperiodic process with purely imaginary frequency $\omega=\mathrm{i}\gamma$:
\begin{equation}
\omega^2=-\gamma^2=(c_\text{A}^2-v_{\text{i}\parallel}^2)\,k_{\parallel}^2=c_\text{A}^2\,(1-\beta_{\parallel}/2)\,k_{\parallel}^2,
\label{SquareOfFrequencyForAlfvenWave}
\end{equation}
where $c_\text{A}^2=B^2/(4\pi m_\text{i}n_{\text{i}\Sigma0})$~--- the square of the Alfven velocity, $n_{\text{i}\Sigma0}=2n_\text{i}^{(1)}$~--- the total ion density in the two streams. The dependence (\ref{SquareOfFrequencyForAlfvenWave}) of the increment $\gamma$ only on the longitudinal component $k_{\parallel}$ of the wave vector means that an arbitrary transverse profile of the current structure grows at the identical rate.

Magnetic sound coincides with the Alfven wave in longitudinal propagation, so it simultaneously undergoes firehose instability. However, for oblique magnetic sound, the plasma parameter $\beta_{\parallel}$ required for instability increases as $2/{\cos^2\theta}$, where $\theta$~--- the angle between the wave vector and the external magnetic field \cite[Chap.~3, \S~10]{Krall-book-eng}. The observed variation in plasma density predominantly across rather than along the external magnetic field corresponds to a superposition of initial perturbations with wave vectors directed at angle $|\theta-\pi/2|\ll1$. Such magnetosonic perturbations remain stable. 

In turn, ionic sound is stable to the anisotropy of ionic temperature and keeps the direction of its group velocity strictly along the external magnetic field (at frequencies below the ionic cyclotron frequency). As a longitudinal electrostatic wave, ionic sound is susceptible to beam instability. The latter transforms the counter mono-velocity flows into a homogeneous distribution of ions along the longitudinal momentum of plateau type, preserving the longitudinal temperature of ions and, consequently, the temperature anisotropy.

\subsection*{4.~SELF-CONSISTENT STATIONARY CURRENT STRUCTURE}

In a homogeneous collisionless plasma with anisotropic temperature, a stationary periodic sequence of current layers {\cite{Veselovsky-1975-transl}, \cite[\S~2.4]{Kocharovsky-2016-transl}, self-supported by its own magnetic field, is possible. In~the limit of small relative perturbation of the system, the distance between plasma density maxima (and magnetic induction nodes)
\begin{equation}
\Delta L_\text{stat}=
\frac{\pi c}{\omega_\text{pi}}\,
\biggl(\frac{T_{\text{i}\parallel}}{T_{\text{i}\perp}}-1\biggr)^{-1/2}\,
\biggl(1+\frac{ZT_{\text{e}\perp}}{T_{\text{i}\perp}}\biggr)^{-1/2}
\label{DistanceBetweenPlasmaDensityMaximaInStationaryStructure}
\end{equation}
coincides with the half period of the wave for which the increment of the Weibel instability becomes zero for anisotropic ions and electrons \cite[Chap.~9, \S~10.2]{Krall-book-eng}. In~expression (\ref{DistanceBetweenPlasmaDensityMaximaInStationaryStructure}) $\omega_\text{pi}$~--- ion plasma frequency; $T_{\text{i}\parallel}$ and $T_{\text{i}\perp}$~--- ion temperature along and across the anisotropy axis ($T_{\text{i}\parallel}>T_{\text{i}\perp}$). The ``transverse'' and ``longitudinal'' electron temperatures are nearly the same ($T_{\text{e}\perp}\approx T_{\text{e}\parallel}$) and are adjusted so that the electric charge density of the plasma turns to zero everywhere. This current system is not~destroyed if an external magnetic field is applied along the higher temperature axis \cite[\S~2.6]{Kocharovsky-2016-transl}. 

If ions are distributed isotropically in velocity at the bases of the experimental magnetic arch, then at the top of the loop the ion temperature ratio $T_{\text{i}\parallel}/T_{\text{i}\perp}$ coincides with the mirror ratio $\alpha$ in the collisionless regime and the distance (\ref{DistanceBetweenPlasmaDensityMaximaInStationaryStructure}) is rewritten in an equivalent form
\begin{multline}
\Delta L_\text{stat}=
\frac{\pi c}{\omega_\text{pi\,foot}}\,
\biggl(1-\frac{1}{\alpha}\biggr)^{-1/2}\,
\biggl(1+\frac{\alpha}{M_\text{s}^2}\biggr)^{-1/2}
\approx\\\approx
3{.}1\,
\biggl(\frac{n_\text{i\,foot}}{1{.}6\cdot10^{15}~\text{cm}^{-3}}\biggr)^{-1/2}\,
\biggl(\frac{1-\alpha^{-1}}{8/9}\biggr)^{-1/2}\,
\biggl(\frac{1+\alpha/M_\text{s}^2}{1{.}7}\biggr)^{-1/2}~\text{cm}.
\label{DistanceBetweenPlasmaDensityMaximaInArcRoof}
\end{multline}
Here, the ionic plasma frequency $\omega_\text{pi}$ of two streams at the top of the system is expressed through the same value at the base of the loop $\omega_\text{pi\,foot}=\sqrt{\alpha}\,\omega_\text{pi}$; the ionic density of one flow $n_\text{i\,foot}$ at the chamber inlet is normalized to the maximum achievable density in the discharge (\ref{MaximalIonDensityInOneFlowAtArcFoot}). The other normalization numbers in the formula (\ref{DistanceBetweenPlasmaDensityMaximaInArcRoof}) are given for the mirror ratio $\alpha=9$, Mach number $M_\text{s}=3{.}5$, charge number $Z=1{.}7$, and mass $m_\text{i}=27~\text{u}$.

The value of (\ref{DistanceBetweenPlasmaDensityMaximaInArcRoof}) is close to the diameter $3~\text{cm}$ of the cord at the top of the tube and, therefore, coincides with the spatial scale of layering in the experiment. This coincidence also indirectly points to the firehose instability as a possible stratification mechanism for the system, since in the~limit of plasma and magnetic pressures of the same order ($\beta_{\parallel}\sim2$) the ionic Weibel instability is transformed into a firehose instability. For the specified threshold ratio $\beta_{\parallel}=2$, the distance under discussion (\ref{DistanceBetweenPlasmaDensityMaximaInStationaryStructure}) turns out to be of the order of the diameter of the Larmor circle of ions $2v_{T\text{i}\perp}/\omega_{B\text{i}}$ with thermal velocity $v_{T\text{i}\perp}=\sqrt{T_{\text{i}\perp}/m_\text{i}}\,$:
\begin{equation*}
\frac{\Delta L_\text{stat}}{2v_{T\text{i}\perp}/\omega_{B\text{i}}}\approx\frac{\pi}{2\,\sqrt{(1-\alpha^{-1})\,(1+\alpha/M_\text{s}^2)}}\approx1{.}26.
\end{equation*}
In~its turn, the zero increment (\ref{SquareOfFrequencyForAlfvenWave}) of the Alfven wave for a plasma density perturbation homogeneous along the magnetic field ($k_{\parallel}=0$) with its arbitrary transverse profile extends the class of stationary current configurations: It does not~limit them to a selected step (\ref{DistanceBetweenPlasmaDensityMaximaInStationaryStructure}), but assumes only an appropriate spatial scale larger than the ionic Larmor radius (for the applicability of the magnetohydrodynamic approach).

\subsection*{5.~TORSIONAL ALFVEN OSCILLATION IN A LONGITUDINALLY INHOMOGENEOUS PLASMA TUBE}

Under the assumption of transversely homogeneous ion density at the arc discharge cathode (as the boundary condition of the problem), the plasma cord stratification originates from the most unstable Alfven-type mode, which grows faster than the system existence time and further saturates at the nonlinear stage (due to successive isotropization of ions, decrease of thermal velocity $v_{\text{i}\parallel}$, and zeroing of the increment (\ref{SquareOfFrequencyForAlfvenWave})). However, Alfven motion almost does not~perturb the density of the medium if the characteristic increment is below the ion cyclotron frequency. Therefore, the firehose instability must develop very rapidly (at the limit of applicability of magnetic hydrodynamics) to create the observed inhomogeneity of ion density.

Based on the energy principle \cite{Bernstein-1958}, perturbation of the magnetic field outside the plasma cord requires energy expenditure of the ion flux, so the most unstable oscillation should be expected to be the torsional Alfven oscillation. In the latter, there is no radial perturbation of the magnetic field, which allows the oscillation to be completely localized inside the plasma cord (without leak into the vacuum). 

\paragraph*{5.1.~Vector potential of the torsional mode.}

Consider the model of a straight (without bending) axially symmetric longitudinally inhomogeneous plasma cord. Let the unperturbed magnetic induction $\mathbf{B}_0=\mathop{\mathrm{rot}}\mathbf{A}_0$ be described by the factorized solenodal vector potential $\mathbf{A}_0=\rho B_{0z}(z)\boldsymbol{\phi}^0/2$ and contains only components along the symmetry axis $\mathbf{z}^0$ and in the radial direction $\boldsymbol{\rho}^0$ of the cylindrical coordinate system $\rho$, $\phi$, $z$:
\begin{equation*}
\mathbf{B}_0=B_{0z}(z)\mathbf{z}^0-\frac{\rho}{2}\,\frac{\mathrm{d}B_{0z}}{\mathrm{d}z}\,\boldsymbol{\rho}^0=B_0(z,\rho)\mathbf{b}^0.
\end{equation*}
Here $\mathbf{b}^0=\mathbf{B}_0/|\mathbf{B}_0|$~--- the tangent vector to the magnetic line of force.

In a torsional Alfven oscillation, the perturbation of magnetic induction $\tilde{\mathbf{B}}=\mathop{\mathrm{rot}}\tilde{\mathbf{A}}$ and electric intensity $\tilde{\mathbf{E}}=-c^{-1}\,\partial\tilde{\mathbf{A}}/\partial t$ is characterized by the vector potential $\tilde{\mathbf{A}}=\rho\tilde{\Phi}(z,\rho,t)B_0\,[\boldsymbol{\phi}^0,\mathbf{b}^0]$ directed along the normal to the magnetic surfaces:
\begin{gather}
\tilde{\mathbf{B}}=\boldsymbol{\phi}^0\mathop{\mathrm{div}}(\rho\tilde{\Phi} B_0\mathbf{b}^0)
-\rho\tilde{\Phi} B_0\,(\boldsymbol{\phi}^0\,\nabla)\,\mathbf{b}^0
=\rho B_0\,(\nabla\tilde{\Phi},\mathbf{b}^0)\,\boldsymbol{\phi}^0,
\label{MagneticInductionDisturbanceInTorsionalAlfvenOscillation}
\\
\tilde{\mathbf{E}}=-\frac{\rho B_0}{c}\,\frac{\partial\tilde{\Phi}}{\partial t}\,[\boldsymbol{\phi}^0,\mathbf{b}^0].
\label{ElectricIntensityDisturbanceInTorsionalAlfvenOscillation}
\end{gather}
In the expression (\ref{MagneticInductionDisturbanceInTorsionalAlfvenOscillation}), the nonzero directional derivative $(\boldsymbol{\phi}^0\,\nabla)\,\mathbf{b}^0=b^0_\rho\boldsymbol{\phi}^0/\rho$ for the radial component $b^0_\rho\boldsymbol{\rho}^0$ of the tangent vector $\mathbf{b}^0$ is taken into account. The value of $\tilde{\Phi}$ is determined with accuracy to the stationary summand $\tilde{\Phi}_0(\pi\rho^2B_{0z})$, which is constant on each magnetic force tube, but allows its dependence on the magnetic flux $\pi\rho^2B_{0z}$, ``numbering'' the specified axially symmetric magnetic surfaces. The difference $\tilde{\Phi}[z',\rho\,\sqrt{B_{0z}(-z_\text{c})/B_{0z}(z')}\,,t]-\tilde{\Phi}(-z_\text{c},\rho, t)$ is equal to the distinction between the azimuth displacement $\phi$ of the perturbed magnetic force line in sections $z=z'$ and $z=-z_\text{c}$, for example, at the cathode.

In~turn, the zero tangential electric field (\ref{ElectricIntensityDisturbanceInTorsionalAlfvenOscillation}) at the metal cathode establishes at the latter a stationary radial profile of the value $\tilde{\Phi}$, and consequently, the azimuthal twist of the magnetic force line between the sections $z=z_\text{c}$ and $z=-z_\text{c}$ with the specified electrodes. The noted confinement of the magnetic field in the metallic tube bases and freedom in the choice of the summand $\tilde{\Phi}_0(\pi\rho^2B_{0z})$ allow us to impose a boundary condition in the form of a stationary opposite magnitude of the value $\tilde{\Phi}$ at the cathodes fixed by the initial perturbation of the induction $\tilde{\mathbf{B}}(z,\rho,t_0)$:
\begin{gather}
\tilde{\Phi}(z_\text{c},\rho,t)=-\tilde{\Phi}(-z_\text{c},\rho,t)=\tilde{\Phi}_\text{c}(\rho);
\nonumber
\\
\tilde{\Phi}_\text{c}(\rho)=
\frac{1}{\rho}
\int\limits_{-z_\text{c}}^{z_\text{c}}
\frac{\tilde{B}_\phi[z',\rho\,\sqrt{B_{0z}(z_\text{c})/B_{0z}(z')}\,,t_0]\,\mathrm{d}z'}
{\sqrt{B_{0z}(z_\text{c})B_{0z}(z')}}\,,
\label{BoundaryConditionOnPhiAtCathodes}
\end{gather}
where the integral takes into account the coupling of the length element
\begin{equation*} \mathrm{d}\ell'=B_0[z',\rho\,\sqrt{B_{0z}(z_\text{c})/B_{0z}(z')}\,]\,\mathrm{d}z'/B_{0z}(z')
\end{equation*}
along the magnetic line of force and the displacement $\mathrm{d}z$ along the applicate~$z$.

\paragraph*{5.2.~Quasi-kinetic description of ions.}

In the considered case of high temperature anisotropy of ions, we neglect the velocity of cyclotron rotation of the particle. Then in the initial magnetic field $\mathbf{B}_0$ the ion moves along its field line with a constant absolute velocity $\mathbf{v}_0=v_{\parallel}\mathbf{b}_0$, and also makes a bending drift in azimuth with velocity $\mathbf{v}_\text{curv}=v_{\parallel}^2\*\mathinner{[\mathbf{b}^0,(\mathbf{b}^0,\nabla)\,\mathbf{b}^0]/\omega_{B\text{i}}}$, which we neglect further. The electromagnetic field (\ref{MagneticInductionDisturbanceInTorsionalAlfvenOscillation}), (\ref{ElectricIntensityDisturbanceInTorsionalAlfvenOscillation}) perturbs the above motion by means of electric drift and variation of the local direction of magnetic induction, which is expressed as the addition of the azimuthal component to the ionic velocity
\begin{equation}
\tilde{\mathbf{v}}^{(1)}=
c\,\frac{[\tilde{\mathbf{E}},\mathbf{B}_0]}{B_0^2}
+v_{\parallel}\,\frac{\tilde{\mathbf{B}}}{B_0}
=\rho\,\biggl[
\frac{\partial\tilde{\Phi}}{\partial t}
+v_{\parallel}\,(\mathbf{b}^0,\nabla\tilde{\Phi})
\biggr]\,\boldsymbol{\phi}^0.
\label{FirstDriftVelocity}
\end{equation}

The subsequent correction $\tilde{\mathbf{v}}^{(2)}$ to the velocity $\mathbf{v}_0$ (in the asymptotic series on the parameter $1/B_0$) describes the drift along the normal $[\boldsymbol{\phi}^0, \mathbf{b}^0]$, for which the Lorentz force $Ze\,[\tilde{\mathbf{v}}^{(2)},\mathbf{B}_0]/c$ provides the necessary rate of change of momentum $m_\text{i}\, \mathrm{d}\tilde{\mathbf{v}}^{(1)}/\mathrm{d}t$ of the drift motion (\ref{FirstDriftVelocity}), and also the correction to the rate of change of velocity $\mathbf{v}_0$ due to the deviation of the particle's trajectory from the unperturbed magnetic force line due to the same drift (\ref{FirstDriftVelocity}):
\begin{gather}
\frac{\mathrm{d}\tilde{\mathbf{v}}^{(1)}}{\mathrm{d}t}
+(\tilde{\mathbf{v}}^{(1)},\nabla)\,v_{\parallel}\mathbf{b}^0
=\frac{Ze}{m_\text{i}c}\,[\tilde{\mathbf{v}}^{(2)},\mathbf{B}_0],
\nonumber
\\
\tilde{\mathbf{v}}^{(2)}=
-\frac{[\boldsymbol{\phi}^0,\mathbf{b}^0]}{\omega_{B\text{i}}\rho}\,
\biggl[
\frac{\partial}{\partial t}
+v_{\parallel}\,(\mathbf{b}^0,\nabla)
\biggr]\,\biggl\{\rho^2\,
\biggl[
\frac{\partial}{\partial t}
+v_{\parallel}\,(\mathbf{b}^0,\nabla)
\biggr]
\tilde{\Phi}\biggr\}.
\label{SecondDriftVelocity}
\end{gather}

Thus, the field of the torsional mode (\ref{MagneticInductionDisturbanceInTorsionalAlfvenOscillation}), (\ref{ElectricIntensityDisturbanceInTorsionalAlfvenOscillation}) does not~create a tangential force component to the unperturbed magnetic line in the linear approximation in magnitude~$\tilde{\Phi}$. Therefore, let us integrate the collisionless kinetic equation for ions over the transverse component $\mathbf{v}_\perp=\mathbf{v}-\mathbf{b}^0\,(\mathbf{b}^0,\mathbf{v})$ of their velocity $\mathbf{v}$, which leads to the continuity equation
\begin{equation}
\frac{\partial f_{\text{i}\parallel}}{\partial t}
+\mathop{\mathrm{div}}\limits_{\mathbf{r}}\bigl[
(v_\parallel\mathbf{b}^0+\tilde{\mathbf{v}}^{(2)})\,f_{\text{i}\parallel}\bigr]=0
\label{ContinuityEquationForIonDistributionOverLongitudinalVelocity}
\end{equation}
for the ion distribution function 
\begin{equation*}
f_{\text{i}\parallel}(v_\parallel,z,\rho,t)=\iint\limits_{|\mathbf{v}_\perp|<\infty} f_\text{i}(\mathbf{v},z,\rho,t)\,\mathrm{d}^2\mathbf{v}_\perp
\end{equation*}
over the conserved velocity component $v_{\parallel}=(\mathbf{b}^0,\mathbf{v})$. Here $f_\text{i}(\mathbf{v},z,\rho, t)$~--- ion distribution function over the three-dimensional velocity $\mathbf{v}$ normalized by the particle density $n_{\text{i}\Sigma}(\mathbf{r})=\iiint_{|\mathbf{v}|<\infty}f_\text{i}(\mathbf{v}, \mathbf{r},t)\,\mathrm{d}^3\mathbf{v}$.

The equation (\ref{ContinuityEquationForIonDistributionOverLongitudinalVelocity}) can be integrated along the paraxial magnetic force line in a linear approximation by the value of $\tilde{\Phi}$ and a homogeneous unperturbed ion distribution function on the cathodes:
\begin{equation}
f_{\text{i}\parallel}(v_\parallel,z,\rho,t)=
\frac{B_0(z,\rho)}{B_0[z_\text{in},\rho\,\sqrt{B_{0z}(z)/B_{0z}(z_\text{in})}\,]}\,
f_\text{in}(|v_\parallel|)
\,
\biggl\{
1+
\frac{1}{\omega_{B\text{i}}}
\mathop{\mathrm{div}}\limits_{\mathbf{r}}\,
[\tilde{\mathbf{v}}^{(1)},\mathbf{b}^0]\bigg|_{z_\text{in},t_\text{in}}^{z,t}
\biggr\}.
\label{NonLocalSolutionForIonVelocityDistribution}
\end{equation}
The paraxial approximation in the solution (\ref{NonLocalSolutionForIonVelocityDistribution}) assumes a small diameter $2\rho$ of the magnetic tube compared to its radius of curvature $1/|{\mathop{\mathrm{rot}}\mathbf{b}^0}|=1/|(\mathbf{b}^0, \nabla)\,\mathbf{b}^0|$ in the sagittal section $\phi=\text{const}$ (the tube is locally approximated by a contiguous conical surface).
In the~formula (\ref{NonLocalSolutionForIonVelocityDistribution}), the cross section $z_\text{in}=-z_\text{c}\mathop{\mathrm{sign}}v_\parallel$ denotes the cathode from which the ion exited with velocity $v_{\parallel}$ ($\mathop{\mathrm{sign}}a$~--- the sign of the number $a$); $f_\text{in}(|v_\parallel|)$~--- transversely uniform distribution of cathode-emitted ions; the vertical line with indices $z_\text{in},t_\text{in}$ and $z,t$ denotes the difference between the divergence at the ion location (in section $z$) at time $t$ and at the cathode $z_\text{in}$ during emission $t_\text{in}=t-(z-z_\text{in})/v_{\parallel}$. If the specified time $t_\text{in}$ happens to be in the past with respect to the initial time $t_0$ in the problem, then the initial location of the ion $z_0=z-v_{\parallel}\,(t-t_0)$ should be considered the point of ``emission'', and the function $f_\text{in}$~--- the initial ion distribution .

\paragraph*{5.3.~Wave equation for torsional oscillation.}

We continue the study of the problem in the asymptotic limit of rapidly increasing or oscillating Alfven modes over the interval of the ion travel time between the cathodes $2\,|z_\text{c}/v_{\parallel}|$, to neglect the nonlocal and noninstantaneous dependence of the solution (\ref{NonLocalSolutionForIonVelocityDistribution}) on the wave magnetic field (\ref{MagneticInductionDisturbanceInTorsionalAlfvenOscillation}) at the cathodes.
The even-velocity $v_{\parallel}$ distribution function of emitted ions $f_\text{in}(|v_\parallel|)$ excludes cross space-time derivatives in the mean drift velocity (\ref{SecondDriftVelocity}) and, consequently, in the component of the ion current density $\mathbf{j}_\text{i}$ normal to the magnetic tube surface:
\begin{equation}
(\mathbf{j}_\text{i},[\boldsymbol{\phi}^0,\mathbf{b}^0])=
-\frac{n_{\text{i}\Sigma0}m_\text{i}c}{B_0}\,
\biggl\{
\rho\,\frac{\partial^2\tilde{\Phi}}{\partial t^2}
+\frac{v_{\text{i}T\parallel}^2}{\rho}\,
(\mathbf{b}^0,\nabla)\,\bigl[\rho^2\,(\mathbf{b}^0,\nabla)\,\tilde{\Phi}\bigr]
\biggr\},
\label{NormalIonCurrentDensity}
\end{equation}
where $v_{\text{i}T\parallel}^2=\int_0^\infty v_{\parallel}^2f_\text{in}(|v_\parallel|)\,\mathrm{d}v_{\parallel}\big/\!\int_0^\infty f_\text{in}(|v_\parallel|)\,\mathrm{d}v_{\parallel}$~--- the square of the ``thermal'' spread in the velocity of ions emitted into the magnetic tube; 
\begin{equation*}
n_{\text{i}\Sigma0}(\mathbf{r})=\frac{B_0(\mathbf{r})}{B_0(z_\text{c})}\int\limits_{-\infty}^{+\infty} f_\text{in}(|v_{\parallel}|)\,\mathrm{d}v_{\parallel}
\end{equation*}
--- unperturbed ion density. 

In~turn, the longitudinal component of the ion current density is due to perturbation of only the magnetic field, but not the electric field (\ref{MagneticInductionDisturbanceInTorsionalAlfvenOscillation}), (\ref{ElectricIntensityDisturbanceInTorsionalAlfvenOscillation}):
\begin{equation}
(\mathbf{j}_\text{i},\mathbf{b}^0)\mathop{=}\limits^\text{def}
Ze\int\limits_{-\infty}^{+\infty}v_{\parallel}f_{\text{i}\parallel}(v_{\parallel},z,\rho,t)\,\mathrm{d}v_{\parallel}
=
\frac{n_{\text{i}\Sigma0}m_\text{i}c}{B_0}\,
v_{\text{i}T\parallel}^2\,
\mathop{\mathrm{div}}\limits_{\mathbf{r}}
\bigl\{
[\boldsymbol{\phi}^0,\mathbf{b}^0]\,\rho\,(\mathbf{b}^0,\nabla)\,\tilde{\Phi}\bigr\}.
\label{LongitudinalIonCurrentDensity}
\end{equation}
Parts of the ion current (\ref{NormalIonCurrentDensity}), (\ref{LongitudinalIonCurrentDensity}) from the magnetic field perturbation (\ref{MagneticInductionDisturbanceInTorsionalAlfvenOscillation}) are proportional to $v_{\text{i}T\parallel}^2$ and are represented together as the current $c\mathop{\mathrm{rot}}\tilde{\mathbf{M}}$ of magnetization
\begin{equation}
\tilde{\mathbf{M}}=\frac{v_{\text{i}T\parallel}^2}{4\pi c_\text{A}^2}\,\tilde{\mathbf{B}}
=\frac{\beta_{\parallel}(z)}{2}\,\tilde{\mathbf{B}}.
\label{IonTransverseMagnetization}
\end{equation}
The solenoidality of the magnetization current is consistent with the change in ionic density only due to the polarization current in the electric field (\ref{ElectricIntensityDisturbanceInTorsionalAlfvenOscillation}). The ion density perturbation is derived by integrating the distribution function (\ref{NonLocalSolutionForIonVelocityDistribution}) over the velocity $v_\parallel$:
\begin{equation}
\tilde{n}_{\text{i}\Sigma}=\frac{n_{\text{i}\Sigma0}}{\omega_{B\text{i}}}
\mathop{\mathrm{div}}\limits_{\mathbf{r}}
\biggl([\boldsymbol{\phi}^0,\mathbf{b}^0]\,\rho\,\frac{\partial\tilde{\Phi}}{\partial t}\biggr)
=-\frac{1}{Ze}
\mathop{\mathrm{div}}\limits_{\mathbf{r}}
\biggl(\frac{\omega_\text{pi}^2\tilde{\mathbf{E}}}{4\pi\omega_{B\text{i}}^2}\biggr).
\label{IonNumberDensityChange}
\end{equation}
The last equality in the formula (\ref{IonNumberDensityChange}) takes into account the constant ratio $n_{\text{i}\Sigma0}/B_0$, so that the ion density (\ref{IonNumberDensityChange}) corresponds to the polarization charge of the plasma fraction with dielectric susceptibility $\omega_\text{pi}^2/(4\pi\omega_{B\text{i}}^2)=c^2/(4\pi c_\text{A}^2)\gg1$.

The inertia-free electrons participate together with ions in the electric drift and motion along the perturbed magnetic force line with velocity (\ref{FirstDriftVelocity}) and thereby nullify the azimuthal electric current of the plasma. At the same time, the longitudinal electron current is set such that the theorem on the circulation of magnetic induction (\ref{MagneticInductionDisturbanceInTorsionalAlfvenOscillation}) and magnetization (\ref{IonTransverseMagnetization}) in projection to the $\mathbf{b}^0$ direction is satisfied. In its integral formulation, the above theorem determines the electron current $I_\text{e}(z,\rho)$ through the cross section $z=\text{const}$ of a magnetic surface of radius $\rho$:
\begin{equation}
I_\text{e}(z,\rho)=\frac{c\rho\,(\tilde{B}_\phi-4\pi\tilde{M}_\phi)}{2}
=\frac{c\rho^2B_0}{2}\,
\biggl(
1-\frac{v_{\text{i}T\parallel}^2}{c_\text{A}^2}
\biggr)\,
(\mathbf{b}^0,\nabla\tilde{\Phi}).
\label{IntegralElectronCurrentThroughCrossection}
\end{equation}

The electron current (\ref{IntegralElectronCurrentThroughCrossection}) varies along the magnetic tube so that the electron and ion electric charges change to opposite values. (Maintaining the electroneutrality of the plasma in a torsional oscillation resembles unipolar diffusion in a magnetic field.)  In~turn, the variable part of the ion fraction charge is uniquely related to the density perturbation (\ref{IonNumberDensityChange}). Let us write this electric balance in integral form for the volume bounded by the axially symmetric magnetic surface and its cross sections $z$ and $-z_\text{c}$:
\begin{equation}
I_\text{e}(-z_\text{c},\rho_\text{c})-I_\text{e}[z,\rho_\text{s}(z,\rho_\text{c})]
+\frac{n_{\text{i}\Sigma0}m_\text{i}c}{B_0}
\int\limits_{-z_\text{c}}^z
2\pi\rho_\text{s}^2(z',\rho_\text{c})\,\frac{\partial^2\tilde{\Phi}}{\partial t^2}\bigg|_{\rho'=\rho_\text{s}(z',\rho_\text{c})}\,\mathrm{d}\ell'=0,
\label{PlasmaCurrentBalanceForZeroElectricCharge}
\end{equation}
where the constancy of the ratio $n_{\text{i}\Sigma0}(\mathbf{r})/B_0(\mathbf{r})$ is taken into account, and the conservation of magnetic flux through the cross section of the selected tube defines the radius of its surface $\rho_\text{s}(z,\rho_\text{c})=\rho_\text{c}\,\sqrt{B_{0z}(-z_\text{c})/B_{0z}(z)}\,;$ $\mathrm{d}\ell'=B_0[z',\rho_\text{s}(z',\rho_\text{c})]\,\mathrm{d}z'/B_{0z}(z')$~--- element of the length of the magnetic line of force on the surface of the tube. 

We differentiate the balance (\ref{PlasmaCurrentBalanceForZeroElectricCharge}) on the applicate $z$ considering the expression (\ref{IntegralElectronCurrentThroughCrossection}) for the current $I_\text{e}$ and obtain the dynamic equation for the value $\tilde{\Phi}$ on each magnetic surface:
\begin{equation}
\frac{B_0}{B_{0z}}\,
\frac{1}{c_\text{A}^2}\,
\frac{\partial^2\tilde{\Phi}}{\partial t^2}=
(\mathbf{b}^0,\nabla)\,
\biggl[
\biggl(1-\frac{v_{\text{i}T\parallel}^2}{c_\text{A}^2}\biggr)\,
\frac{B_0}{B_{0z}}\,
(\mathbf{b}^0,\nabla)\,\tilde{\Phi}
\biggr]
.
\label{GeneralDynamicEquationOnAzimuthalDisplacement}
\end{equation}
The equation (\ref{GeneralDynamicEquationOnAzimuthalDisplacement}) can also be derived as a theorem on the magnetic field circulation in projection on the normal $[\boldsymbol{\phi}^0,\mathbf{b}^0]$ to the magnetic surface, as well as the Euler equation in ideal magnetic hydrodynamics with strongly anisotropic ionic pressure. The outlined kinetic approach is necessary in that it provides: a)~an ion density perturbation (\ref{IonNumberDensityChange}) to interpret the plasma stratification in the experiment; b)~an expression (\ref{IntegralElectronCurrentThroughCrossection}) for the electron current to formulate the correct boundary condition at special cross sections $z=\pm z_\text{hose}$ where the coefficient at the senior spatial derivative in equation (\ref{GeneralDynamicEquationOnAzimuthalDisplacement}) goes to zero~--- at the boundary of the subapical region of the firehose instability.

The asymptotic linear wave equation (\ref{GeneralDynamicEquationOnAzimuthalDisplacement}) still permits itself to be treated as a Cauchy problem with some initial distribution of $\tilde{\Phi}(z, \rho,t_0)$ and its derivative $\partial\tilde{\Phi}/\partial t$ (which define the initial electromagnetic field (\ref{MagneticInductionDisturbanceInTorsionalAlfvenOscillation}), (\ref{ElectricIntensityDisturbanceInTorsionalAlfvenOscillation})). This approach assumes that the initial ion density in the tube is approximately matched to the cathode emission: $|\tilde{n}_{\text{i}\Sigma}(z,\rho,t_0)|\ll n_{\text{i}\Sigma0}$. The boundary condition on the cathodes is defined by the formula (\ref{BoundaryConditionOnPhiAtCathodes}).

The spatial derivative $(\mathbf{b}^0,\nabla)$ only along the magnetic force line in the equation (\ref{GeneralDynamicEquationOnAzimuthalDisplacement}) means independent dynamics of twisting of different magnetic surfaces inside the plasma cord. This property occurs because the longitudinal current of inertial-free electrons (\ref{IntegralElectronCurrentThroughCrossection}) adjusts to an arbitrary radial profile of the torsion angle $\tilde{\Phi}(\rho)$.

\paragraph*{5.4.~Condition on the faces of the firehose instability region.}

In~the low plasma pressure limit ($\beta_{\parallel}=2v_{\text{i}T\parallel}^2/c_\text{A}^2\ll1$), the equation (\ref{GeneralDynamicEquationOnAzimuthalDisplacement}) describes the stable torsional oscillations of the magnetic tube enclosed in the resonator between the cathodes. As the ion pressure increases, the firehose instability turns on in the central section of the magnetic tube $z=0$ with the weakest magnetic field and further occupies the layer between sections $z=\pm z_\text{hose}$ in which the plasma parameter $\beta_{\parallel}$ takes a threshold value of $2$. In the planes $z=\pm z_\text{hose}$, the coefficient at the higher spatial derivative in the differential equation (\ref{GeneralDynamicEquationOnAzimuthalDisplacement}) goes to zero, so in their vicinity the value $\tilde{\Phi}$ and its spatial derivative can sharply increase in absolute value \cite[\S~10.10]{Richtmyer-book-eng}. This circumstance requires a special boundary condition on the quantity $\tilde{\Phi}$ in the vicinity of cross sections $z=\pm z_\text{hose}$ in addition to the condition (\ref{BoundaryConditionOnPhiAtCathodes}) on the cathodes.

Let us select the $|z\pm z_\text{hose}|\le\Delta z_\text{MHD}/2$ layers where the drift approximation for the ion trajectory and, hence, the wave equation (\ref{GeneralDynamicEquationOnAzimuthalDisplacement}) are violated. Their thickness 
\begin{equation}
\Delta z_\text{MHD}=v_{\text{i}T\parallel}/\omega_{B\text{i}}
\label{MinimalDistanceForMegnetohydrodynamics}
\end{equation}
of the order of the distance traveled by an ion over a quarter of a cyclotron period. The first (``standard'') condition at the special points $z=\pm z_\text{hose}$ comes from the continuity of the electric intensity (\ref{ElectricIntensityDisturbanceInTorsionalAlfvenOscillation}), and hence the continuity of the value $\tilde{\Phi}$ in the narrow layer $|z\pm z_\text{hose}|\le\Delta z_\text{MHD}/2$:
\begin{gather}
\tilde{\Phi}(-z_\text{hose}-\Delta z_\text{MHD}/2,\rho,t)=\tilde{\Phi}(-z_\text{hose}+\Delta z_\text{MHD}/2,\rho,t),
\nonumber
\\
\tilde{\Phi}(z_\text{hose}-\Delta z_\text{MHD}/2,\rho,t)=\tilde{\Phi}(z_\text{hose}+\Delta z_\text{MHD}/2,\rho,t).
\label{ContinuityOfPhiAtSpecialPoints}
\end{gather}

A similar requirement for magnetic induction (\ref{MagneticInductionDisturbanceInTorsionalAlfvenOscillation}) in the form of continuity of the spatial derivative $(\mathbf{b}^0,\nabla\tilde{\Phi})$ is in contradiction with the continuity of the electron current (\ref{IntegralElectronCurrentThroughCrossection}), since the derivative $(\mathbf{b}^0,\nabla\tilde{\Phi})$ is allowed to grow as $1/|z-z_\text{hose}|$ when approaching the cross section $z=z_\text{hose}$, whereas the difference $1-v_{\text{i}T\parallel}^2/c_\text{A}^2$ in the expression (\ref{IntegralElectronCurrentThroughCrossection}) is proportional to $|z-z_\text{hose}|$ and changes sign at cross section $z=z_\text{hose}$. In this approach, the non-zero electron current (\ref{IntegralElectronCurrentThroughCrossection}) is directed in opposite directions from the cross sections $z_\text{hose}\pm\Delta z_\text{MHD}$, which corresponds to the third-party source of particles in the layer $|z-z_\text{hose}|<\Delta z_\text{MHD}$.

The matching of magnetic induction (\ref{MagneticInductionDisturbanceInTorsionalAlfvenOscillation}) and magnetization (\ref{IonTransverseMagnetization}) with the multiplier $4\pi$ in the cross sections $z=\pm z_\text{hose}$ corresponds to the infinite magnetic permeability of the ionic component (for the Alfven wave polarization of the field) with its value changing from $+\infty$ to $-\infty$ (when moving from the outside to the inside of the subapical region of the firehose instability). The magnetic field of the longitudinal electron current $2I_\text{e}/(c\rho)$, acting as a third-party source for ion magnetization, can be taken as the magnetic strength $\mathbf{H}$. Therefore, at special points $z=\pm z_\text{hose}$, the continuity of the electron current (\ref{IntegralElectronCurrentThroughCrossection}) must be required, whereas the solenoidal ion magnetization current $c\mathop{\mathrm{rot}}\tilde{\mathbf{M}}$ generates a current layer along the cross sections $z=\pm z_\text{hose}$ for its self-looping, as on the domain wall of an antiferromagnet. The above circumstance is expressed not in the coincidence but in the oppositeness of the derivative $(\mathbf{b}^0,\nabla\tilde{\Phi})$ on different sides of the cross sections $z=\pm z_\text{hose}$:
\begin{gather}
\Delta z_\text{MHD}\,(\mathbf{b}^0,\nabla\tilde{\Phi})|_{z=-z_\text{hose}-\Delta z_\text{MHD}/2}=-\Delta z_\text{MHD}\,(\mathbf{b}^0,\nabla\tilde{\Phi})|_{z=-z_\text{hose}+\Delta z_\text{MHD}/2},
\nonumber
\\
\Delta z_\text{MHD}\,(\mathbf{b}^0,\nabla\tilde{\Phi})|_{z=z_\text{hose}-\Delta z_\text{MHD}/2}=-\Delta z_\text{MHD}\,(\mathbf{b}^0,\nabla\tilde{\Phi})|_{z=z_\text{hose}+\Delta z_\text{MHD}/2}.
\label{ChangeOfLongitudinalDerivativeOfPhiAtSpecialPoints}
\end{gather}
The paper \cite{Gubchenko-2015} first drew attention to the infinite magnetic permeability of the plasma in a wave with half period (\ref{DistanceBetweenPlasmaDensityMaximaInStationaryStructure})~--- threshold with respect to the Weibel instability.

\paragraph*{5.5.~Stationary part of the current structure.}

The solution of the linear equation (\ref{GeneralDynamicEquationOnAzimuthalDisplacement}) admits a stationary part, which simultaneously satisfies the stationary boundary condition on the cathodes (\ref{BoundaryConditionOnPhiAtCathodes}):
\begin{equation}
\tilde{\Phi}(z,\rho,t)=
\tilde{\Phi}_\text{c}[\varrho(z,\rho)]
\,
\frac{\int_0^z\mathrm{d}z'/(1-v_{\text{i}T\parallel}^2/c_\text{A}^2)}
{\int_0^{z_\text{c}}\mathrm{d}z'/(1-v_{\text{i}T\parallel}^2/c_\text{A}^2)}\,.
\label{StationaryPartOfPhi}
\end{equation}
In the~expression (\ref{StationaryPartOfPhi}), when taking integrals from the function $1/\{1-v_{\text{i}T\parallel}^2/c_\text{A}^2[z',\varrho(z',\rho)]\}$, the layers $|z'\pm z_\text{hose}|\le\Delta z_\text{MHD}/2$ should be excluded, which is equivalent to computing the integral in the sense of the principal value. The function $\tilde{\Phi}_\text{c}(\varrho)$ is defined in the formula (\ref{BoundaryConditionOnPhiAtCathodes}), and the normalized radial variable $\varrho(z,\rho)=\rho\,\sqrt{B_{0z}(z)/B_{0z}(z_\text{c})}$ ``numbers'' the magnetic surfaces and coincides with the radius $\rho_\text{c}$ of the latter at the cathode.

In the stationary solution (\ref{StationaryPartOfPhi}), there is no ionic polarization current in the electric field (\ref{ElectricIntensityDisturbanceInTorsionalAlfvenOscillation}), and hence no perturbation of the ionic density (\ref{IonNumberDensityChange}). The electron current (\ref{IntegralElectronCurrentThroughCrossection}) is found to be homogeneous for each magnetic surface, and its radial profile is proportional to the function $B_{0z}(z)\rho^2\tilde{\Phi}_\text{c}[\varrho(z,\rho)]/2$. In~particular, the electron current can go to zero for the entire cross section of the plasma cord, as in a coaxial cable with the core and braid shorted at the ends. The distribution (\ref{StationaryPartOfPhi}) of the quantity $\tilde{\Phi}$ describes ``transverse'' magnetization of the ion fraction by a stationary electron current (self-circuited through cathodes) with a sharp variation of the azimuthal component of magnetic induction (\ref{MagneticInductionDisturbanceInTorsionalAlfvenOscillation}) near the cross sections $z=\pm z_\text{hose}$, as if on a domain wall in an antiferromagnet.

\paragraph*{5.6.~Exponentially increasing and oscillating perturbations of ion density.}

The plasma density perturbation observed in the experiment is generated only by a nonstationary solution of the equation (\ref{GeneralDynamicEquationOnAzimuthalDisplacement}). Therefore, we consider harmonic and exponentially increasing solutions of the equation (\ref{GeneralDynamicEquationOnAzimuthalDisplacement}) under firehose instability. 

For analytical consideration of the problem we choose a model spatial profile of Alfven velocity 
\begin{equation}
c_\text{A}=c_\text{A\,top}\mathop{\mathrm{ch}}\biggl(\frac{4z}{L}\biggr),
\label{ModelSpatialProfileOfAlfvenVelocity}
\end{equation}
which reflects the minimum of magnetic induction at the top of the loop and its increase toward the bases of the arch at a distance of the order of a quarter of the tube length $L/4$ ($c_\text{A\,top}$~--- Alfven velocity at the top of the arch $z=0$). Then the equation (\ref{GeneralDynamicEquationOnAzimuthalDisplacement}) is instantiated to a dimensionless form
\begin{equation}
\frac{\partial^2\tilde{\Phi}}{\partial\tau^2}=
\frac{\partial}{\partial\eta}
\biggl\{
\biggl[1-\frac{\beta_{\parallel\text{top}}}{2}\,(1-\eta^2)\biggr]\,(1-\eta^2)\,
\frac{\partial\tilde{\Phi}}{\partial\eta}
\biggr\}.
\label{DimensionlessModelDynamicEquation}
\end{equation}
Here the time variable 
\begin{equation}
\tau=\frac{4c_\text{A\,top}t}{L}
\label{DimensionlessTimeDefinition}
\end{equation}
is normalized by the Alfven wave travel time of the quarter arch. The longitudinal spatial variable $\eta(z)=\mathop{\mathrm{th}}(4z/L)$ varies in the interval $-1<\eta<1$; $\beta_{\parallel\text{top}}=2\,v_{\text{i}T\parallel}^2/c_\text{A\,top}^2$~--- ratio of longitudinal ionic and magnetic pressure at the top of the arch. The partial derivatives on the variable $\eta$ are computed at a fixed normalized radius $\varrho$, which is defined after formula (\ref{StationaryPartOfPhi}).

The cathode positions in the boundary condition (\ref{BoundaryConditionOnPhiAtCathodes}) correspond to points $|\eta|=\eta_\text{c}=\sqrt{1-\alpha^{-1}}$ where the magnetic induction exceeds the same value at the top of the arch by approximately the mirror ratio $\alpha\gg1$. In~turn, the boundary conditions (\ref{ContinuityOfPhiAtSpecialPoints}), (\ref{ChangeOfLongitudinalDerivativeOfPhiAtSpecialPoints}) occur at the special points $|\eta|=\eta_\text{hose}=\sqrt{1-2/\beta_{\parallel\text{top}}}$ where the ``transverse'' ionic magnetic permeability goes to infinity and changes its sign.

The solution of the initial problem for the equation (\ref{DimensionlessModelDynamicEquation}) in the region $|\eta|<\eta_\text{hose}$ is, strictly speaking, an incorrectly posed problem similar to Adamar's example in the case of the Cauchy problem for the Laplace equation \cite[\S~1.4.8]{Vladimirov-book-transl}. The incorrectness is removed by introducing a viscosity contributing dissipation of short-wave perturbations \cite{Wong-2005}. We limit our search to perturbations with exponential time dependence $\exp(\pm\varkappa\tau)$. The differential operator in the right-hand side of the equation (\ref{DimensionlessModelDynamicEquation}) is characterized by a sign-variable coefficient at the second spatial derivative, so the corresponding Sturm~--- Liouville problem should be referred to the right-defined \cite{Mingarelli-2011, Richardson-1918}: the sign of the Lagrangian can be either positive or negative, but the extrema of the Lagrangian are defined at a fixed sign-defined functional as the norm of the function. Our goal is to prove a significant increase of the perturbation most extended along the magnetic tube during the time of existence of the experimental system (of the order of two times of the ion's passage of the length of the arch). 

Higher plasma pressure at the top of the arch increases the rate of growth of the perturbation. Therefore, it is sufficient to consider the case of a small excess of $\beta_{\parallel\text{top}}-2\ll1$ of the parameter $\beta_{\parallel\text{top}}$ over the threshold value $2$, when the cross sections $|\eta|=\eta_\text{hose}\ll1$ of the zero ``twist'' elasticity of the magnetic field are far away from the cathodes $|\eta|=\eta_\text{c}\approx1$. In~the above limit, the equation (\ref{DimensionlessModelDynamicEquation}) for perturbations of the type $\tilde{\Phi}(z,\rho,t)=\exp[\pm\varkappa\tau(t)]\tilde{\Phi}_\varkappa[\tilde{\eta}(z),\varrho(z,\rho)]$ coincides with the Legendre equation
\begin{equation}
\frac{\mathrm{\partial}}{\mathrm{\partial}\tilde{\eta}}
\,(1-\tilde{\eta}^2)\,
\frac{\mathrm{\partial}\tilde{\Phi}_\varkappa}{\mathrm{\partial}\tilde{\eta}}
+\varkappa^2\tilde{\Phi}_\varkappa=0,
\label{LegendreEquation}
\end{equation}
where the spatial variable $\tilde{\eta}=\eta/\eta_\text{hose}$ is normalized by the ``distance'' $\eta_\text{hose}$ from the apex of the arch to the boundaries of the instability region and allows its model variation from $-\infty$ to ${+\infty}$. 

Boundary condition (\ref{BoundaryConditionOnPhiAtCathodes}) of zero electric intensity at cathode points \begin{equation*}
\tilde{\eta}_\text{c}=\frac{\sqrt{1-\alpha^{-1}}}
{\sqrt{1-2/\beta_{\parallel\text{top}}}}\gg1
\end{equation*}
is detailed for stationary ($\varkappa=0$) and nonstationary ($\varkappa\ne0$) components:
\begin{equation}
\tilde{\Phi}_\varkappa(\tilde{\eta}_\text{c},\varrho)=-\tilde{\Phi}_\varkappa(-\tilde{\eta}_\text{c},\varrho)=
\begin{cases}
\tilde{\Phi}_\text{c}(\varrho),&\varkappa=0;\\
0,&\varkappa\ne0.
\end{cases}
\label{DetailedBoundaryConditionOnPhiAtCathodesForStationaryAndHarmonicParts}
\end{equation}

The general solution of the equation (\ref{LegendreEquation}) is represented by a factorized linear combination
\begin{equation}
\tilde{\Phi}_\varkappa(\tilde{\eta},\varrho)=
a^{(\text{e})}_\varkappa(\varrho)w^{(\text{e})}_{\nu(\varkappa)}(\tilde{\eta})
+a^{(\text{o})}_\varkappa(\varrho)w^{(\text{o})}_{\nu(\varkappa)}(\tilde{\eta})
\label{LinearCombinationAsSolutionOfLegendreEquation}
\end{equation}
of even ($w^{(\text{e})}_\nu$) and odd ($w^{(\text{o})}_\nu$) functions on the variable $\tilde{\eta}$
\begin{multline}
w^{(\text{e})}_\nu(\tilde{\eta})=
\frac{P_\nu(|\tilde{\eta}|)+\mathop{\mathrm{Re}}P_\nu(-|\tilde{\eta}|)}
{2\mathop{\mathrm{Re}}\cos(\pi\nu/2)}
=\\
\shoveright{=
\begin{cases}
\displaystyle
\cos\Bigl(\frac{\pi\nu}{2}\Bigr)P_\nu(|\tilde{\eta}|)
-\frac{2}{\pi}\sin\Bigl(\frac{\pi\nu}{2}\Bigr)\mathop{\mathrm{Re}}Q_\nu(|\tilde{\eta}|),
&\mathop{\mathrm{Re}}\nu\ge-\frac{1}{2}\,, \mathop{\mathrm{Im}}\nu=0;\\
\displaystyle
\frac{P_\nu(|\tilde{\eta}|)
+(2/\pi)\mathop{\mathrm{ch}}(\pi\mathop{\mathrm{Im}}\nu)\mathop{\mathrm{Re}}Q_\nu(|\tilde{\eta}|)}
{\sqrt{2}\mathop{\mathrm{ch}}(\pi\mathop{\mathrm{Im}}\nu/2)}\,,
&\mathop{\mathrm{Re}}\nu=-\frac{1}{2}\,, \mathop{\mathrm{Im}}\nu\ge0,
\end{cases}
}\\
\shoveleft{
w^{(\text{o})}_\nu(\tilde{\eta})=\mathop{\mathrm{sign}}(\tilde{\eta})\,
\frac{P_\nu(|\tilde{\eta}|)-\mathop{\mathrm{Re}}P_\nu(-|\tilde{\eta}|)}
{2\mathop{\mathrm{Re}}\sin(\pi\nu/2)}
=}\\
=\mathop{\mathrm{sign}}(\tilde{\eta})\times
\begin{cases}
\displaystyle
\sin\Bigl(\frac{\pi\nu}{2}\Bigr)P_\nu(|\tilde{\eta}|)
+\frac{2}{\pi}\cos\Bigl(\frac{\pi\nu}{2}\Bigr)\mathop{\mathrm{Re}}Q_\nu(|\tilde{\eta}|),
&\mathop{\mathrm{Re}}\nu\ge-\frac{1}{2}\,, \mathop{\mathrm{Im}}\nu=0;\\
\displaystyle
\frac{-P_\nu(|\tilde{\eta}|)
+(2/\pi)\mathop{\mathrm{ch}}(\pi\mathop{\mathrm{Im}}\nu)\mathop{\mathrm{Re}}Q_\nu(|\tilde{\eta}|)}
{\sqrt{2}\mathop{\mathrm{ch}}(\pi\mathop{\mathrm{Im}}\nu/2)}\,,
&\mathop{\mathrm{Re}}\nu=-\frac{1}{2}\,, \mathop{\mathrm{Im}}\nu\ge0.
\end{cases}
\label{EvenAndOddSolutionsOfLegendreEquation}
\end{multline}
Here, the modified Legendre function of the first kind of zero order $P_\nu(x)$ \cite[\S~14]{NIST-Handbook} regularly passes the special point $x=1$ and takes purely real values in the interval $-1<x<+\infty$ for complex degree 
\begin{equation}
\nu(\varkappa)=\bigl(-1+\sqrt{1+4\varkappa^2}\,\bigr)/2,
\label{DegreeOfLegendreFunctionViaIncrement}
\end{equation}
in particular, $P_\nu(1)=1$. 

The analytic continuation of the function $P_\nu(x)$ into the complex plane of argument $x$ \cite[formula~(14.24.1)]{NIST-Handbook} takes complex-conjugate values at the upper and lower edges of the cut $-\infty<x<-1$, and hence exhibits the same real part
\begin{equation*}
\mathop{\mathrm{Re}}P_\nu(-|\tilde{\eta}|)=
\mathop{\mathrm{Re}}\bigl[\cos(\pi\nu)P_\nu(|\tilde{\eta}|)-(2/\pi)\sin(\pi\nu)Q_\nu(|\tilde{\eta}|)\bigr].
\end{equation*}
The modified Legendre function of the second kind of zero order $Q_\nu(x)$ takes real values in the interval $1<x<+\infty$ only in the case of the real degree $\nu$, but its analytic continuation to the upper or lower edge of the cut ${-\infty}<x<1$ of the complex plane $x$ shows the same real part on the interval $-1<x<1$ with a logarithmic divergence
\begin{equation}
\mathop{\mathrm{Re}}Q_\nu(x)=\frac{1}{2}\ln\biggl(\frac{2}{|x-1|}\biggr)
+\psi(1)-\mathop{\mathrm{Re}}\psi(\nu+1)
\label{LogarithmicDivergenceOfLegendreQNearUnitArgument}
\end{equation}
at $x\to1\pm0$ for any degree (\ref{DegreeOfLegendreFunctionViaIncrement}); $\psi(y)=\Gamma'(y)/\Gamma(y)$~--- digamma function, or the logarithmic derivative of the gamma function $\Gamma(y)$ \cite[formula~(5.2.2)]{NIST-Handbook}.

The condition of continuity (\ref{ChangeOfLongitudinalDerivativeOfPhiAtSpecialPoints}) for electron current (\ref{IntegralElectronCurrentThroughCrossection}) at the special points $\tilde{\eta}=\pm1$ requires equality of coefficients before the logarithmically divergent part of the type (\ref{LogarithmicDivergenceOfLegendreQNearUnitArgument}) in the solution (\ref{LinearCombinationAsSolutionOfLegendreEquation}) of the equation (\ref{LegendreEquation}) (on each magnetic surface) on opposite sides from the specified points.\,\footnote{The Liouville transformation \cite{Karjanto-2024} translates the Legendre equation (\ref{LegendreEquation}) into the Schrödinger equation for a particle of constant mass in a one-dimensional potential well $U[\chi(\tilde{\eta})]$ with singularities of the form $U\propto1/[\chi\mp\chi(1)]^2$ along the independent coordinate $\chi(\tilde{\eta})$. Such features of the potential correspond to the boundary case of a particle falling on the attracting center \cite[\S~18, 35]{Landau-III-transl}. At the same time, under the above transformation, the unit coefficient at the eigenvalue $\nu\,(\nu+1)$ changes sign to the opposite value in the Schrödinger equation on leaving the segment of the independent variable $|\chi|<\chi(1)$, which is characteristic of the left-defined Sturm~--- Liouville problem. As a result, the classical rotation points appear to be distant from the special points $\pm\chi(1)$ by less than the wavelength of the oscillation and are qualitatively similar to the variant of the paper \cite{Ishkhanyan-2017-transl}.} Functions (\ref{EvenAndOddSolutionsOfLegendreEquation}) share this property. 

In~its turn, the continuity (\ref{ContinuityOfPhiAtSpecialPoints}) of the electric intensity (\ref{ElectricIntensityDisturbanceInTorsionalAlfvenOscillation}) sets the same coefficients in front of the regular part in the solution on opposite sides of the singular point $\tilde{\eta}=\pm1$ on the background of slow logarithmic divergence (\ref{LogarithmicDivergenceOfLegendreQNearUnitArgument}) on the distance $||\tilde{\eta}|-1|$. Thus, the basis functions (\ref{EvenAndOddSolutionsOfLegendreEquation}) satisfy the conditions (\ref{ContinuityOfPhiAtSpecialPoints}) and (\ref{ChangeOfLongitudinalDerivativeOfPhiAtSpecialPoints}) at the boundary of the firehose instability region.

As a result, the condition (\ref{DetailedBoundaryConditionOnPhiAtCathodesForStationaryAndHarmonicParts}) of zero electric intensity at the cathode becomes the dispersion equation ($w^{(\text{e})}_\nu(\eta_\text{c})=0$ or $w^{(\text{o})}_\nu(\eta_\text{c})=0$) for the dimensionless increment $\varkappa=\sqrt{\nu\, (\nu+1)}$ in the case of positive degree $\nu$ or frequency 
\begin{equation*}
\varpi=-\mathrm{i}\varkappa=
\begin{cases}
\sqrt{-\nu\,(\nu+1)}\,,&-1/2\le\nu<0;\\
\sqrt{(1/4)+(\mathop{\mathrm{Im}}\nu)^2}\,,
&\mathop{\mathrm{Re}}\nu=-1/2,\mathop{\mathrm{Im}}\nu\ge0
\end{cases}
\end{equation*}
of torsional oscillation:
\begin{gather}
\mathop{\mathrm{ctg}}\Bigl(\frac{\pi\nu^{(\text{e})}}{2}\Bigr)=
\frac{2Q_{\nu^{(\text{e})}}(\tilde{\eta}_\text{c})}
{\pi P_{\nu^{(\text{e})}}(\tilde{\eta}_\text{c})}\,,
\qquad
-\mathop{\mathrm{tg}}\Bigl(\frac{\pi\nu^{(\text{o})}}{2}\Bigr)=
\frac{2Q_{\nu^{(\text{o})}}(\tilde{\eta}_\text{c})}
{\pi P_{\nu^{(\text{o})}}(\tilde{\eta}_\text{c})}\,;
\label{DispersionEquationOnDimensionlessIncrement}
\\
\frac{1}{\mathop{\mathrm{ch}}(\pi\upsilon^{(\text{e})})}=
-\frac{2\mathop{\mathrm{Re}}Q_{-1/2+\mathrm{i}\upsilon^{(\text{e})}}(\tilde{\eta}_\text{c})}
{\pi P_{-1/2+\mathrm{i}\upsilon^{(\text{e})}}(\tilde{\eta}_\text{c})}\,,
\qquad
\frac{1}{\mathop{\mathrm{ch}}(\pi\upsilon^{(\text{o})})}=
\frac{2\mathop{\mathrm{Re}}Q_{-1/2+\mathrm{i}\upsilon^{(\text{o})}}(\tilde{\eta}_\text{c})}
{\pi P_{-1/2+\mathrm{i}\upsilon^{(\text{o})}}(\tilde{\eta}_\text{c})}\,.
\label{DispersionEquationOnDimensionlessFrequency}
\end{gather}
The degree $\nu$ in~equation (\ref{DispersionEquationOnDimensionlessIncrement}) is real, and in equation (\ref{DispersionEquationOnDimensionlessFrequency})~--- is complex and is denoted as $\nu=-1/2+\mathrm{i}\upsilon$, $\upsilon\ge0$. 

The right-hand sides of the equations (\ref{DispersionEquationOnDimensionlessIncrement}) are positive and monotonically decreasing on the variable $\nu$ as
\begin{equation*}
\frac{2Q_\nu(\tilde{\eta}_\text{c})}
{\pi P_\nu(\tilde{\eta}_\text{c})}
\approx
\frac{2\,[\Gamma(\nu+1)]^2}
{(2\tilde{\eta}_\text{c})^{2\nu+1}\,\Gamma(\nu+1/2)\Gamma(\nu+3/2)}
\end{equation*}
in the considered limit of distant cathodes $\tilde{\eta}_\text{c}\gg1$ according to their asymptotics \cite[formulas~(14.8.12), (14.8.15), (14.3.10)]{NIST-Handbook}. As a~result, the spectrum of the real degree $\nu$ is close to the integers:
\begin{equation}
\nu_k\approx (k-1)-\frac{4\,[\Gamma(k)]^2}
{\pi\,(2\tilde{\eta}_\text{c})^{2k-1}\,\Gamma(k-1/2)\Gamma(k+1/2)}\,,
\label{SpectrumOfLegendreFunctionDegreeForIncrement}
\end{equation}
where the natural index $k=1,2,3\ldots$ coincides with the number of cross sections $z=\text{const}$ between the cathodes in which the value of $\tilde{\Phi}_\varkappa$, and hence the electric field strength (\ref{ElectricIntensityDisturbanceInTorsionalAlfvenOscillation}) and the density perturbation (\ref{IonNumberDensityChange}) go to zero. The solutions (\ref{EvenAndOddSolutionsOfLegendreEquation}) almost coincide with the irregular function $Q_\nu(\tilde{\eta})$ except in the near-cathode region, where the regular component $P_\nu(\tilde{\eta})$ rapidly decaying from the cathodes provides zero electric intensity at these electrodes.

In particular, the solution with one node of value $\tilde{\Phi}_{\sqrt{\nu_1\,(\nu_1+1)}}$ ($k=1$) is close in longitudinal spatial structure to the stationary solution (\ref{StationaryPartOfPhi}) (on each magnetic surface). Its negative degree (\ref{SpectrumOfLegendreFunctionDegreeForIncrement}) corresponds to the real frequency $\omega_1\approx8c_\text{A\,top}/(\pi L\,\sqrt{\tilde{\eta}_\text{c}}\, )$ that decreases with relative separation of the cathodes $\tilde{\eta}_\text{c}$ (narrowing of the subapical region of the firehose instability). This circumstance indicates the possibility of slow beats of the field of the stationary solution (\ref{StationaryPartOfPhi}) and the lowest-frequency nonstationary part (with index $k=1$) with frequency $\omega_1$. The odd electric intensity profile (\ref{ElectricIntensityDisturbanceInTorsionalAlfvenOscillation}) and, consequently, the ion density perturbation (\ref{IonNumberDensityChange}) do not~allow us to consider this mode responsible for the observed plasma stratification (which is even).

The quasi-cylindrical plasma stratification most extended along the arch, even along the $z$ coordinate, is associated with the $k=2$ mode ($\nu_2\approx1$). The increment of this mode 
\begin{equation}
\gamma_2\approx\frac{4\,\sqrt{2}\,c_\text{A\,top}}{L}
\label{LowerIncrementOfFirehoseInstabilityInMagneticArc}
\end{equation}
is almost independent of the position of the distant cathodes, as well as of the extent of the unstable region, and hence of the excess of the $\beta_{\parallel\text{top}}$ parameter over the~$2$ threshold. The sign of the ion density perturbation is opposite at the apex of the arch and at the boundaries of the firehose instability region in this mode (on the same magnetic surface). Therefore, the observed quasi-cylindrical increase in density on the outer wall of the tube in the subapical region should be accompanied by a similar increase on the tube axis near the ends of the instability region (like bright filamentary sections).

The main spectrum of stable Alfven modes is described by the dispersion equation (\ref{DispersionEquationOnDimensionlessFrequency}) for the complex degree $\nu$ of the form $-1/2+\mathrm{i}\upsilon$ and is bounded from below by the dimensionless frequency $\varpi=-\mathrm{i}\varkappa=1/2$. In the~extremely distant cathode limit ($\ln(2\tilde{\eta}_\text{c})\gg\pi$), the ``low-frequency'' part of this spectrum ($|\upsilon|\ll1$, $-\mathrm{i}\varkappa\approx1/2+\upsilon^2$) corresponds to Alfven modes that tunnel efficiently through the region of the firehose instability. The equations (\ref{DispersionEquationOnDimensionlessFrequency}) take the form \cite[Chap.~5, formulas (143), (152)]{Hobson-1931}
\begin{equation*}
(-1)^\sigma=-\mathop{\mathrm{ctg}}[\upsilon\ln(2\tilde{\eta}_\text{c})]\mathop{\mathrm{sh}}(\pi\upsilon)
\end{equation*}
and give approximately equidistant but different steps in the imaginary part of degree $\nu=-1/2+\mathrm{i}\upsilon$ for even ($\sigma=0$) and odd ($\sigma=1$) modes:
\begin{equation*}
\upsilon_q=\frac{\pi\,(q+1)}{\ln(2\tilde{\eta}_\text{c})}
\biggl[1-(-1)^\sigma\frac{\pi}{\ln(2\tilde{\eta}_\text{c})}\biggr].
\end{equation*}
Here, the integer index $q=0,1,2\ldots$ measures the number $\varsigma$ of nodal cross sections $z=\text{const}$ between the cathodes in the even ($\varsigma^{(\text{e})}=2q$) and odd ($\varsigma^{(\text{o})}=2q+1$) modes for the value~$\tilde{\Phi}$. In the~even mode, all nodes are located in the stable region $|\tilde{\eta}|>1$, and in the odd mode, a node is added in the median section $\tilde{\eta}=0$. The degrees $\upsilon$ follow alternately for even and odd modes: $\upsilon_q^{(\text{e})}<\upsilon_q^{(\text{o})}<\upsilon_{q+1}^{(\text{e})}$.

The ``high-frequency'' part of the spectrum ($\varpi\approx\upsilon\gg1$) corresponds to modes that do not practically~penetrate through the firehose instability layer $|\tilde{\eta}|<1$, so that the oscillations in the regions $\tilde{\eta}<-1$ and $\tilde{\eta}>1$ are mutually independent. The equations (\ref{DispersionEquationOnDimensionlessFrequency}) are transformed to the expression \cite[Chap.~5, formulas (143), (152)]{Hobson-1931}
\begin{equation*}
(-1)^\sigma=-\mathop{\mathrm{ctg}}[\upsilon\ln(2\tilde{\eta}_\text{c})+\pi/4]\mathop{\mathrm{ch}}(\pi\upsilon).
\end{equation*}
Successive frequencies from even and odd modes form duplets
\begin{equation*}
\upsilon_q=\frac{1}{\ln(2\tilde{\eta}_\text{c})}\,
\biggl\{
\pi\,(q+3/4)
-\frac{(-1)^\sigma}{\mathop{\mathrm{ch}}[\pi\,(q+3/4)/{\ln(2\tilde{\eta}_\text{c})]}}
\biggr\},
\end{equation*}
within which the splitting narrows as the number of nodes $q$ in the mode increases in the interval $1<\tilde{\eta}<\tilde{\eta}_\text{c}$. The characteristic step $\pi/\ln(2\tilde{\eta}_\text{c})$ between neighboring frequencies for a mode of one parity shrinks very slowly (logarithmically) with relative cathode separation $2\tilde{\eta}_\text{c}$.

\subsection*{6.~SCENARIO OF FIREHOSE INSTABILITY IN A MAGNETIC ARCH}

The incorrectness of the Cauchy problem for the dynamic equation (\ref{DimensionlessModelDynamicEquation}) comes from the asymptotically linear growth of the increment (\ref{SpectrumOfLegendreFunctionDegreeForIncrement}) with increasing number of nodes $k\gg1$ in the unstable mode (as in Adamar's example for the Laplace equation). However, ion drift motion with velocities (\ref{FirstDriftVelocity}), (\ref{SecondDriftVelocity}) is realized only in modes, in which the nodes and maxima of the electric intensity (\ref{ElectricIntensityDisturbanceInTorsionalAlfvenOscillation}) and magnetic induction (\ref{MagneticInductionDisturbanceInTorsionalAlfvenOscillation}) follow each other along the magnetic surface with a step greater than the distance $\pi\,\Delta z_\text{MHD}$, see expression (\ref{MinimalDistanceForMegnetohydrodynamics}). Shorter-wavelength modes grow in the regime of ionic Weibel instability with cold electrons, the increment of which is different from zero only when the non-zero electron inertia is taken into account \cite[formula~70]{Davidson-1972}, \cite[Chap.~9, \S~10.4]{Krall-book-eng}, \cite{Mikhailovsky-I-book-transl}.

Characteristic number of unstable firehose modes in the magnetic arch
\begin{equation*}
n_\text{max}=
\frac{L\omega_{B\text{i\,top}}\beta_{\parallel\text{top}}
\ln\bigl(\beta_{\parallel\text{top}}/2+\sqrt{\beta_{\parallel\text{top}}/2-1}\,\bigr)}
{4\pi v_{\text{i}T\parallel}}
\end{equation*}
we estimate as the ratio of the unstable region extent
\begin{equation*}
\textstyle
2z_\text{hose}=L\ln\bigl(\beta_{\parallel\text{top}}/2+\sqrt{\beta_{\parallel\text{top}}/2-1}\,\bigr)/2
\end{equation*}
to the specified distance $\pi\,\Delta z_\text{MHD}=2\pi v_{\text{i}T\parallel}/(\omega_{B\text{i\,top}}\beta_{\parallel\text{top}})$ near the cross section $z=z_\text{hose}$ for the model profile (\ref{ModelSpatialProfileOfAlfvenVelocity}) of Alfven velocity. 

For the setup parameters, the close-to-two value of $2\pi v_{\text{i}\parallel}/(\omega_{B\text{i\,top}}L)\approx1{.}8$ for magnetic induction (\ref{MagneticInductionOfObservedPlasmaLayering}) reflects the general limitation of the magnetohydrodynamic approach at the apex of the loop. The first unstable mode ($n_\text{max}=1$) appears at parameter $\beta_{\parallel\text{top}}\approx3{.}7$, which is almost $1{.}5\text{--}2{.}0$ times higher than the threshold value $\beta_{\parallel}=2$ in a homogeneous plasma ($L=\infty$). In~this case, the unstable region $|z|<z_\text{hose}$ extends half the length of the arch: $2z_\text{hose}/L=\ln\bigl(\beta_{\parallel\text{top}}/2+\sqrt{\beta_{\parallel\text{top}}/2-1}\,\bigr)/2\approx0{.}5$.

During the time of increasing field of the unstable mode by a factor of $2{.}7$ with increment (\ref{LowerIncrementOfFirehoseInstabilityInMagneticArc}) the ion with velocity $v_{\text{i}T\parallel}$ travels a distance $L\,\sqrt{\beta_{\parallel\text{top}}}/8\approx0{. }24L$~--- approximately one-quarter of an arch, which allows the firehose instability to enter the nonlinear regime over the lifetime of the system. At the same time, the increment (\ref{LowerIncrementOfFirehoseInstabilityInMagneticArc}) is close to the ion cyclotron frequency at the top of the arch $\omega_{B\text{i\,top}}$:
\begin{equation*}
\frac{\gamma_2}{\omega_{B\text{i\,top}}}=
\frac{\sqrt{2\beta_{\parallel\text{top}}}\ln\bigl(\beta_{\parallel\text{top}}/2+\sqrt{\beta_{\parallel\text{top}}/2-1}\,\bigr)}{\pi n_\text{max}}=0{.}88,
\end{equation*}
---~and thus provides the possibility of significant perturbation of the ion density (\ref{IonNumberDensityChange}).

Thus, the quite compact size of the ``Solar Wind'' setup with respect to the minimum distance (\ref{MinimalDistanceForMegnetohydrodynamics}) of the applicability of magnetic hydrodynamics seems to ensure the existence of only one unstable firehose mode in the arc. Its increment (\ref{LowerIncrementOfFirehoseInstabilityInMagneticArc}) of the order of the ion cyclotron frequency allows the firehose instability to enter the nonlinear regime during the lifetime of the system and perturb the ion density to a detectable level in the form of a near-wall cylindrical layer.

\subsection*{CONCLUSIONS}

The experimental setup ``Solar Wind'' creates a compact plasma cord of the coronal arch type, in which the ion pressure along the external magnetic field of the mirror trap type significantly exceeds the analogous pressure in the transverse direction. When the anisotropic pressure at the top of the experimental arch exceeds the threshold value for the development of firehose instability, the plasma cord is observed to stratify. The present work substantiates the detected layering of the type of a near-wall cylindrical layer of increased plasma density as a manifestation of an unstable torsional Alfven mode. The rather compact size of the setup (not~slightly exceeding the distance traveled by the ion over a cyclotron period) provides a high increment of firehose instability of the most extended along the arch torsional mode, which is able to perturb appreciably the plasma density during the lifetime of the system.

At the linear stage, the unstable plasma motion resembles two adjoining axisymmetric convective cells of firehose turbulence. The central cross section of the loop forms a nominal baffle between the cells, along which ions flow from the axis to the tube wall and form the observed near-wall cylindrical layer. The outer edges of the cells pass along the borders of the region of firehose instability (where the Alfven velocity coincides with the longitudinal ion thermal velocity): here the ions, on the contrary, drift along the radius to the tube axis. The longitudinal electron current short-circuits the above transverse ion currents along the axis and the outer wall of the arch inside the unstable region.

In the considered spatially inhomogeneous system, the boundaries of the firehose instability region are associated with an infinite magnetic permeability of the ionic fraction (with respect to the azimuthal component of the magnetic induction perturbation). As a~result, the longitudinal electron current excites an ionic current layer at the specified boundary, which becomes similar to a domain wall in an antiferromagnet (the nonzero azimuthal perturbation of magnetic induction is opposite in direction on different sides of this wall).

The authors thank V.~M.~Gubchenko for pointing out the infinite magnetic permeability of the unmagnetized plasma at threshold medium parameters for the Weibel instability.

The study is supported by the Russian Science Foundation (project No. 23-12-00317 ``Interaction of supersonic plasma flows in magnetic arch'').

\printbibliography 
\end{document}